# Mapping topographic, geophysical and gravimetry data of Pakistan – a contribution to geological understanding of Sulaiman Fold Belt and Muslim Bagh Ophiolite Complex

*Polina Lemenkova* [1*]

[1)] Université Libre de Bruxelles, École polytechnique de Bruxelles (Brussels Faculty of Engineering), Laboratory of Image Synthesis and Analysis (LISA). Building L, Campus de Solbosch, Avenue Franklin Roosevelt 50, Brussels 1000, Belgium.

*Author's correspondence: polina.lemenkova@ulb.be or pauline.lemenkova@gmail.com

*Abstract*

*Along with the development of the scripting technology in cartography, such as the Generic Mapping Tools (GMT) and libraries of R programming language, geologic and geophysical mapping is being vigorously promoted, where the integration of the thematic data, such as GEBCO/SRTM, EGM-2008 and open geological raster and vector layers is one of the primary datasets that provides the high-resolution raw sources for cartographic visualization in the geologically complex regions like Pakistan. This study aims to integrate scripting methods of automated cartography, methods of applied geoinformatics for geomorphometric analysis and technical data processing (formatting, projecting, plotting), to provide a synthesis of the geological, geophysical and geomorphological maps of Pakistan they as new information supporting analysis of the geospatial variations of geology, geomorphology, tectonics and gravity fields with a special focus on the geologically remarkable region of Pakistan: Sulaiman Fold Belt and Muslim Bagh ophiolite complex. This study presents new 12 thematic maps, which are technically made using scripting approaches and open tools. All maps cover the region of Pakistan and they are made using open source tools: GMT, R and QGIS. A GMT and R based scripting mapping is applied for mapping Pakistan, and its algorithm steps are presented stepwise as code snippets. A system complex approach of the data integration and formats reshaping, data conversion and reformatting for a single project of the geology of Pakistan is designed and developed based on the combination of the programming and scripting techniques and with additional QGIS based mapping, which effectively integrates the thematic geospatial multi-origin datasets. Various color palettes and cartographic visualization approaches have been used to achieve the best visualization. The resulting maps are explained and discussed. Correlation between spatial phenomena of Earth's gravity, geologic evolution and tectonic movements were pointed out and commented on. New 12 maps present the regional geologic setting of the country.*

## 1 Introduction

The complexity in the geologic structure of Pakistan (Fig. 1) and richness of mineral resources (*Kazmi and Jan*, 1997; *Sadiq Malkani*, 2014; *Rehman et al.*, 2016; *Mengal et al.*, 1994) driven by the geologic and tectonic evolution (*Zia et al.*, 2018; *Barry et al.*, 2002; *Ahmed et al.*, 2018) has prompted efforts aimed at geologic and geophysical modelling of the region supported by thematic mapping for better understanding of the geology of the Pakistani region (*Honarmand et al.*, 2020; *Lemenkova*, 2020a; *Inam et al.*, 2019). Modern geomorphic structure of landforms of areas with a highly varied and contrasting terrain relief, such as Pakistani, often mirrors the tectonic development and geodynamic evolution of the region that includes multiple processes, such as faulting, high seismicity, orogenesis, sedimentation processes, etc. (*Gohl et al.*, 2006a, 2006b; *Lemenkova*, 2019a). It is critical that geologically important areas of Pakistan be visualized properly with regular updates of the maps based on the actual high-resolution datasets and modern cartographic tools. This especially refers to such regions as Balochistan and Sulaiman Fold Belt and Muslim Bagh ophiolite complex (*Kerr et al.*, 2016; *Kakar et al.*, 2014; *Khan et al.*, 2007).

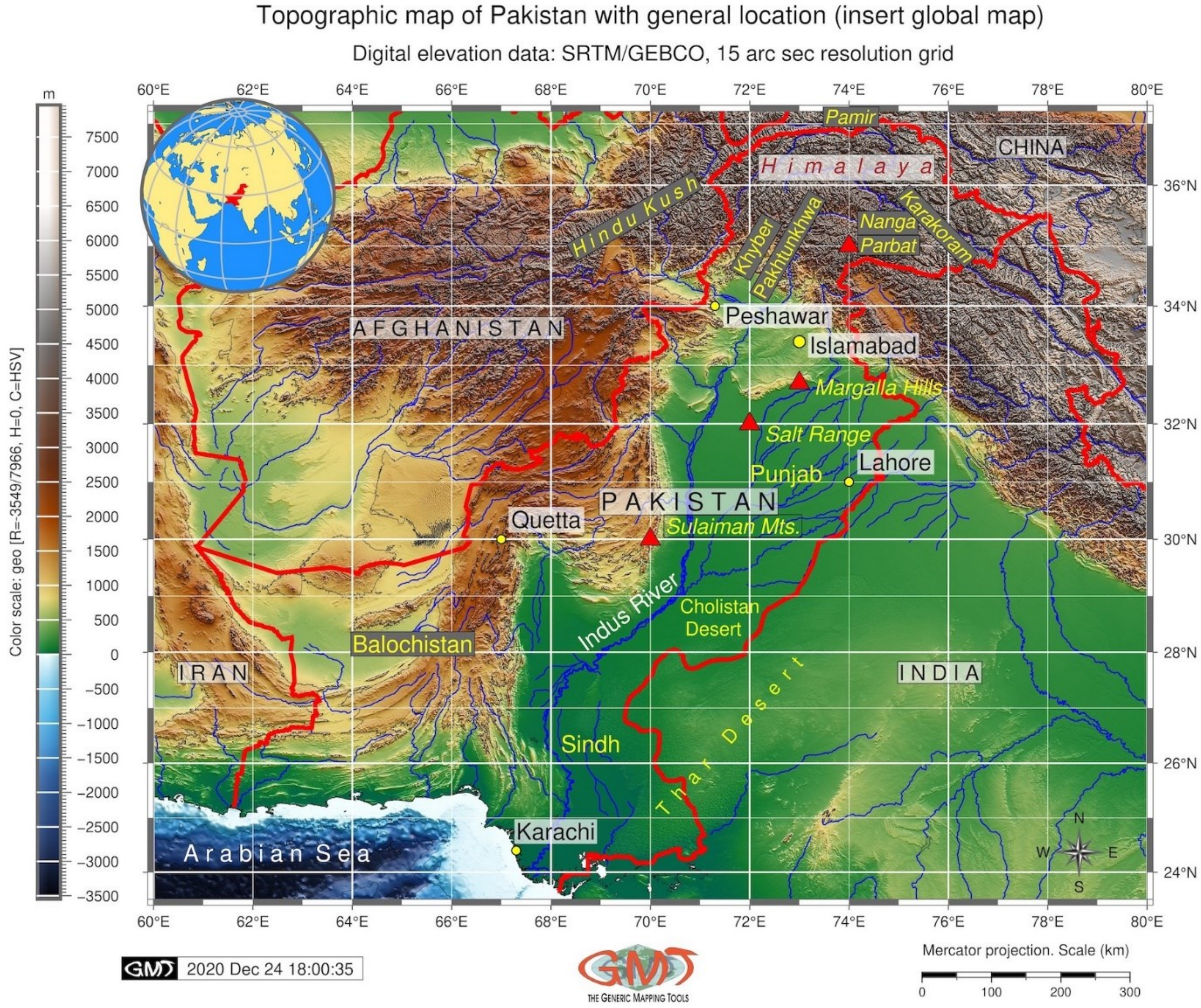

Fig. 1. Topographic map of Pakistan. Mapping: GMT. Data source: GEBCO (*GEBCO Compilation Group*, 2020). Map production: author.

The orogenic regions of Pakistan contain a high number of mineral resources valuable in geologic exploration (*Ahmed et al.*, 2020; *Kakar et al.*, 2013). Proper cartographic visualization optimizes geologic resources for effective prognosis, estimating hazard risks, enables effective mineral prospecting and natural resource monitoring (*Lemenkova et al.*, 2012; *Klaučo et al.*, 2014; *Lemenkova*, 2020b; *Panezai et al.*, 2020). In addition, using scripting approaches in geologic mapping has been found to promote increased cartographic precision, effectiveness, and productivity of topographic, geochemical, petrographic, and geophysical data processing (*Din et al.*, 2019; *Lemenkova*, 2020c) which is caused by the fundamental principle of the programming languages that implies repeatability of loops in a workflow (*Lemenkova*, 2019b, 2019c) which results in automatization of mapping and reduce of faults.

Using free open source geospatial datasets that combine topographic, geologic and geophysical data visualized by the GMT and QGIS, as well as DEM based modelling by R programming language developed a new dataset source on Pakistan relief. This is useful for studying morphological responses of the orogenic relief to plate tectonic movements causing complex geological processes reflected in its current geomorphology. The data collected by means of detailed mapping together with the thematic data (geologic, geophysical and tectonic, sediments) serve as fundamental elements for cartographic visualization of the study area, thus establishing the basis for successful completion of the planned geologic studies (*Kuhn et al.*, 2006).

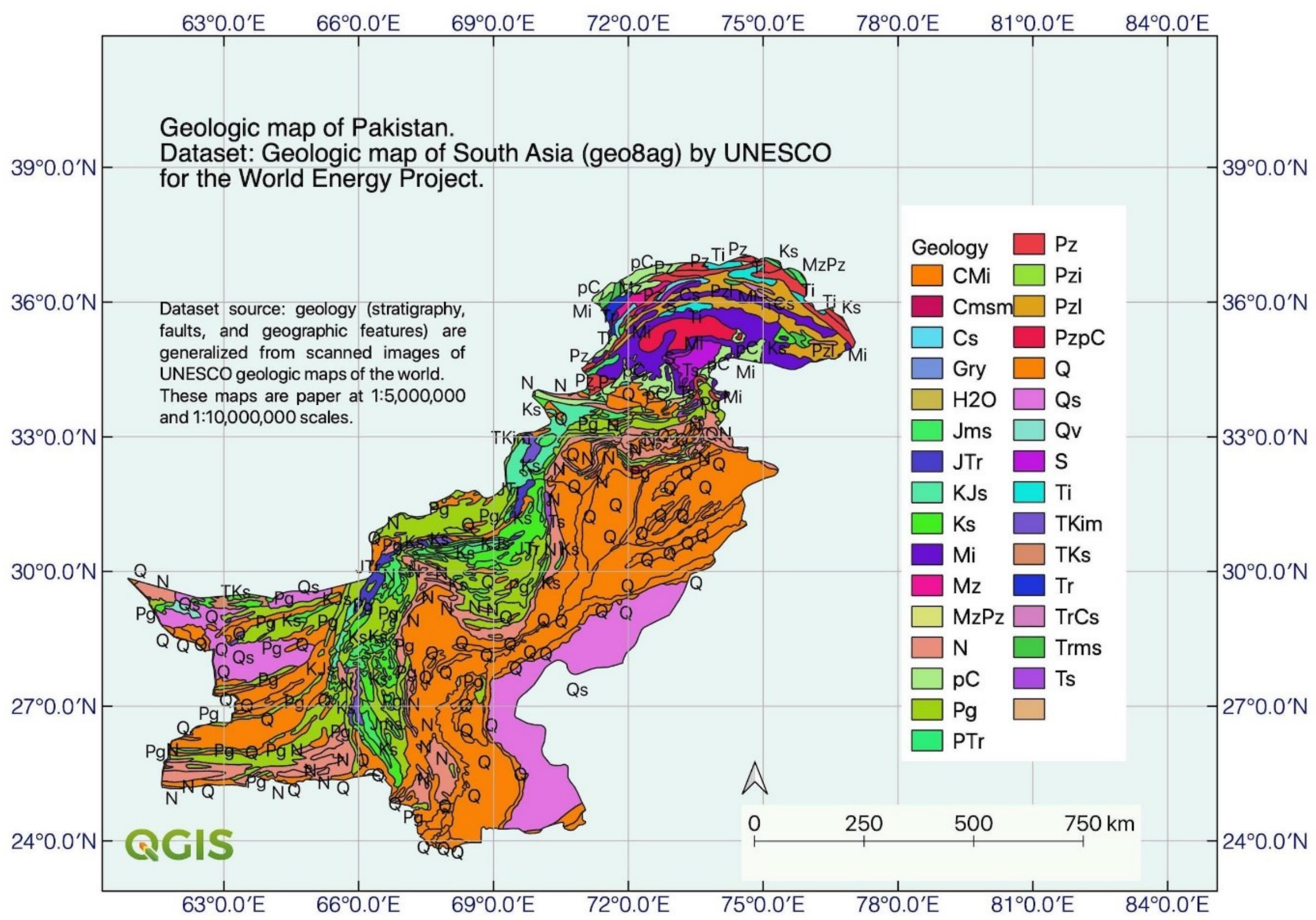

Fig. 2. Lithologic map of Pakistan. Mapping: QGIS. Data source: USGS (*Wandrey and Law*, 1998). Map production: author.

This paper demonstrates the application of the scripting cartographic mapping of Pakistan using Generic Mapping Tools (GMT), R language, and QGIS, supported by the integrated usage of open geospatial datasets. These combine GEBCO/SRTM raster grid for topographic mapping, EGM-2008 grid for geophysical mapping, UNESCO dataset for geological mapping, and R libraries for processing of the DEM-based geomorphometric analysis (slope, aspect, hillshade, and relief map) to give insights into the geomorphology of Pakistan. Much effort has thus been directed to develop methods to effectively and smoothly visualize geological, topographic and geophysical datasets using various cartographic approaches. Among others, tested methods include statistical modelling (*Räisänen*, 2005; *Khan et al.*, 2016; *Klaučo et al.*, 2013a), comparative spatial data analysis (*Aftab et al.*, 2013; *Klaučo et al.*, 2013b) machine learning and automatization (*Lemenkova*, 2019d; *Schenke and Lemenkova*, 2008), 3D modelling (*Akiska et al.*, 2013; *Lemenkova*, 2020d; *Abid et al.*, 2019), cartographic overlays, and mapping layouts for data processing and visualization using Google Earth or GIS in Earth and environmental studies (*Suetova et al.*, 2005a, 2005b; *Nasir and Arslan*, 2012; *Poutanen and Steffen*, 2014; *Kierulf et al.*, 2019). This study contributes to the existing wealth of the research in Earth sciences with a focus on integrated studies of Pakistan.

Integrating multi-disciplinary datasets is needed to better understanding and characterization of the geomorphology and topographic structure of the area (*Lemenkova*, 2021c). However, comprehensive data collection over such a large study area as Pakistan results in relatively large datasets that can be too time-consuming in cartographic workflow and visualization, which arises the question of the scripting applied for processing of such datasets (*Lemenkova*, 2020f). Thus, a modular system of GMT scripting toolset that can synthesize multiple lines of code in a single script that can be repeated for various map layouts has become particularly valuable in research projects where extensive datasets are combined in a thematic project (integrating geological, geophysical, geomorphological and topographic data) which is difficult to process using standard GIS.

GIS-based 2D and 3D mapping supported by graphical plotting of statistical graphs and visualizing remote sensing aerial data have become quite common for geologic mapping of Pakistan (*Siddiqui et al.*, 2016; *Jiskani et al.*, 2018). Other examples of data processing techniques include stratigraphic columns (*Shah*, 1977), statistical data analysis (*Lemenkova*, 2020g). However, advancements in the use of scripting techniques of cartography to integrate machine learning approaches in the geospatial analysis have primarily been confined to reports on cartographic techniques and details of workflow (*Lemenkova*, 2020e, 2021b). Far less attention has been paid to the development of integrated approaches that combines several tools for geological mapping based on multi-format thematic datasets (vector shape files visualized by QGIS, GRD raster grids mapping by GMT, spatial analysis performed by R) to perform a complex visualization of Pakistan. Most existing GIS-based maps of Pakistan are either based on the traditional GIS, such as ArcGIS (*Yasin et al.*, 2020; *Khan*, 2019; *Khan et al.*, 2019; *Kanwal et al.*, 2017), on using special data obtained in fieldwork, e.g., synthetic seismograms, pseudo-seismic sections, surface resistivity logs and sounding (Akhter *et al.*, 2012), or focused on specific questions of the Pakistani geology: fluvial depositional system (*Ullah et al.*, 2009). Thus, although many approaches in a GIS can be applied for geologic mapping, few complete integrated frameworks are available to perform geophysical and geological 2D and 3D modelling and spatial analysis.

This research technically tests the approach of an integrated cartographic framework built from using the combined application of tools (QGIS, GMT, and R programming language) and multi-source datasets (topography, geomorphology, geology, geophysics) to perform a comparative 2D and 3D mapping of the provinces of Pakistan with a special accent on Balochistan, Himalaya suites, Karakorum, and Muslim Bagh ophiolite complex, which aims to contribute the overview of the Pakistani geology and geomorphology for further regional thematic applications in geosciences. The present scripting framework is rapidly applied thanks to the repeatability of codes by GMT and R, regional in scale, based on the free-of-charge datasets, and built on multi-source thematic data covering Pakistan. While the traditional GIS-based mapping often uses commercial software, the advantages of the present research offer an efficient alternative by providing comprehensive methods of mapping using free scripting tools and datasets.

## *2 Materials and Methods*

### *2.1. Data*

The Generic Mapping Tools (GMT) scripts were used to map several illustrations in this paper: Figure 1 (topographic map), Figure 3 (tectonic map), Figures 4, 5, and 6 (geophysical maps of geoid and gravity), Figures 7 and 8 (3D relief models), from open datasets and visualize raster grids for thematic mapping. The automated nature of the GMT (*Wessel et al.*, 2019) cartographic scripting toolset allowed efficient visualization of the high-resolution data on geology, topography, and geophysics across Pakistan.

The first step in the mapping procedure was to select input data as the raw materials. These include the following available datasets: GEBCO grid with 15 arc-second resolution (*GEBCO Compilation Group*, 2020) which is based on the SRTM (*Becker et al.*, 2009) for plotting Fig. 1, geologic data in Fig. 2 utilize the dataset from the USGS (*Wandrey and Law*, 1998), ETOPO1 grid for plotting a map of ophiolites in Fig. 3 (*Amante and Eakins*, 2009), Fig. 4 is based on geoid dataset EGM-2008 (*Pavlis et al.*, 2012), geophysical grids (*Sandwell and Smith*, 1997; *Sandwell et al.*, 2014) used for plotting Figs. 5 and 6, the ETOPO5 used for plotting Figs. 7 and 8 based on (*GDAL/OGR,* 2020) with grid resolution reduced to 5 minutes.

### *2.2. Scripting in GMT*

Next, the data were visualized by the GMT using a set of modules and GMT syntax stepwise using shell scripts. The GMT modules include among the most the following ones: `psbasemap`, `grdimage`, `psscale`, `grdview`, `grdcontour`, `psbasemap`, `pstext`, `img2grd`, `grd2cpt`, to mention some of them. These modules were chosen for their applicability for raster image processing in GRD and IMG formats, visualization of color palettes, cartographic elements, layout map composition, availability of projections for cartographic visualization of data, and ability to flexibly manipulate with graphical elements on the map for plotting geospatial data.

#### *2.2.1. Topographic mapping*

Figures 1 and 3 represent topographic and geologic maps of Pakistan made using a combination of GMT modules. The SRTM data (Shuttle Radar Topography Mission), DEM on a near-global scale from 56°S to 60°N, presents the most complete high-resolution digital topographic cartographic data of Earth, is developed as an international project by the NGA and NASA. The part of the SRTM grid covering Pakistan was a subset in GMT using the following code: `gmt grdcut GEBCO_2019.nc -R60.0/80.0/23.5/37.2 -Gpk_relief.nc`. Here the option `-R60.0/80.0/23.5/37.2` defines the region in WESN coordinate convention.

The image was then visualized by the `grdimage` module using the code: `gmt grdimage pk_relief.nc -Cmyocean.cpt -R60.0/80.0/23.5/37.2 -JM6.5i -I+a15+ne0.75 -Xc -K > $ps`. The isolines on the relief were plotted using the code `gmt grdcontour pk_relief.nc -R -J -C1000 -W0.1p -O -K >> $ps` and the insert map was plotted using the countries codes: `ISO 3166-1 alpha-2`. Here is the code for Pakistan: `gmt pscoast -Rg -JG70/30N/$w -Da -Gpeachpuff -A5000 -Bga -Wfaint -EPK+gred -Slightskyblue1 -O -K -X$x0 -Y$y0 >> $ps`.

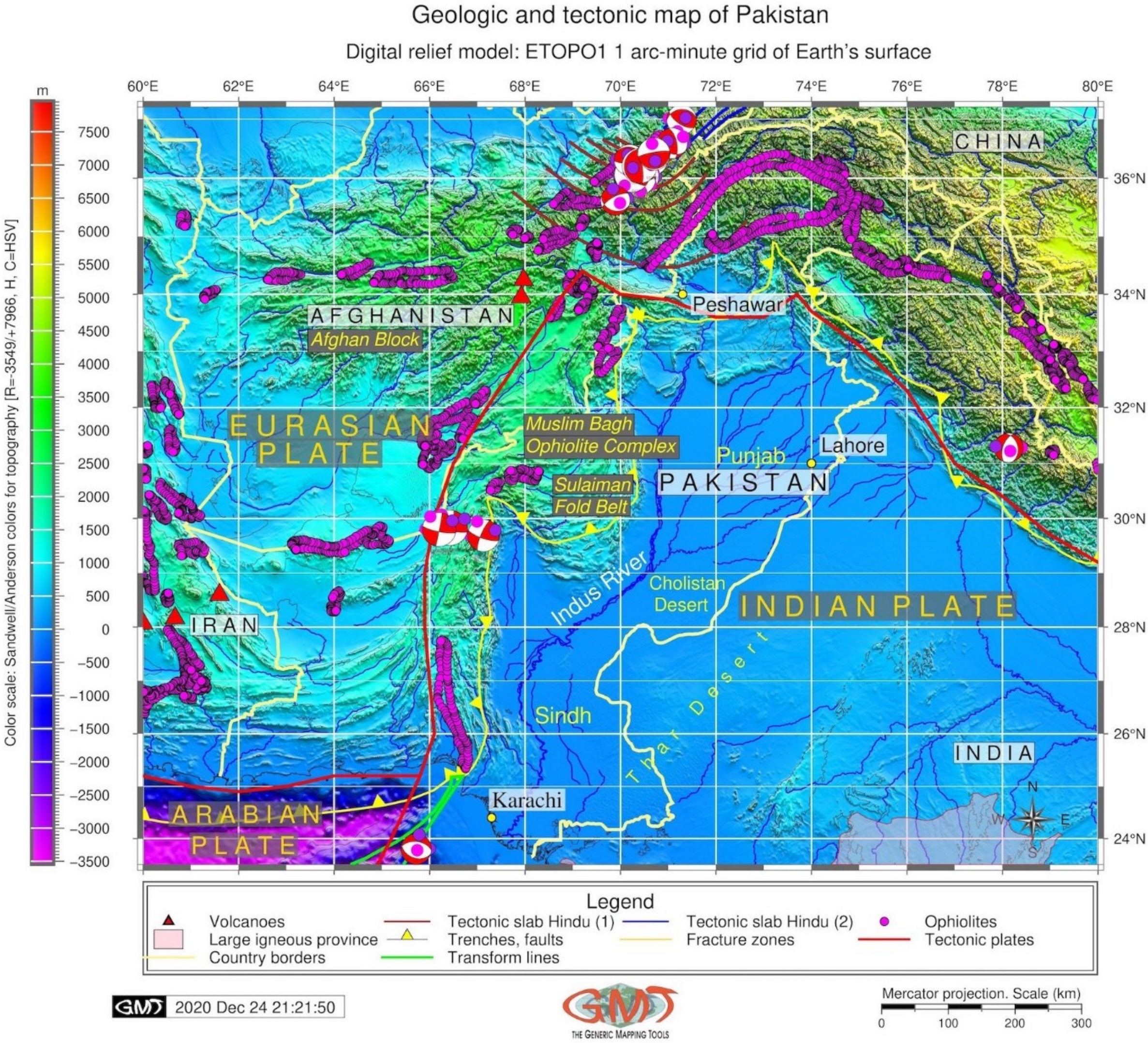

Fig. 3. Geologic map of Pakistan. Mapping: GMT. Data source: ETOPO1 (*Amante and Eakins*, 2009). Map production: author.

### *2.2.2. Geologic mapping*

The tectonic slab contours were plotted using the following codes: `gmt psxy -R -J SC_hindu1.txt -Wthinner,purple -O -K >> $ps` (the same code repeated for `SC_hindu2.txt` and `SC_assam.txt`). The plates were mapped using the code: `gmt psxy -R -J TP_Arabian.txt -L -Wthickest,red -O -K >> $ps` and `gmt psxy -R -J TP_Indian.txt -L -Wthickest,red -O -K >> $ps`. The ophiolites were plotted using the code `gmt psxy -R -J ophiolites.gmt -Sc0.15c -Gmagenta -Wthinnest -O -K >> $ps`. The geologic data set consists of lithological datasets provided by UNESCO which includes the petroleum geology, geologic provinces, and oil and gas fields of South Asia, which is a cartographic part of a worldwide series released by the USGS World Energy Project obtained in shape (`.shp`) file ArcGIS format and processed in the open source QGIS (*QGIS.org*, 2020) and visualized in Fig. 2.

### *2.2.3. Geoid modelling*

The data selected for geoid (Fig. 4) were initially stored in an `.adf` file, which is one component of an Esri GRID file. Therefore, the data were converted to the GMT GRD format using the following code: `gmt grdconvert n00e45/w001001.adf geoid_PK.grd`.

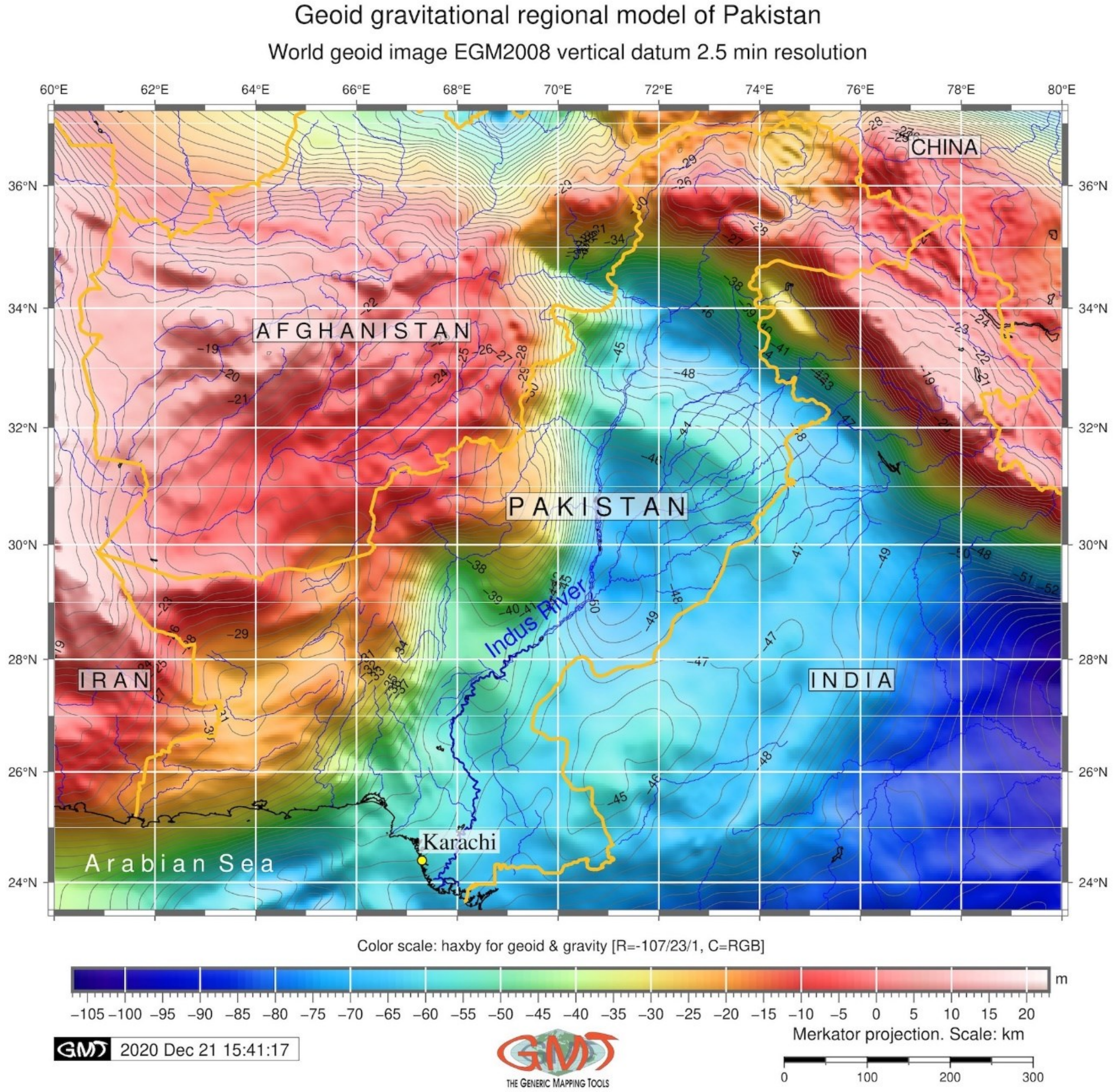

Fig. 4. Geoid model of Pakistan. Mapping: GMT. Data source: EGM-2008 (*Pavlis et al.,* 2012). Map production: author.

The information (data range) was then checked up using `gdalinfo` utility of GDAL as follows: `gdalinfo geoid_PK.grd -stats`. The color palette was defined using the code: `gmt makecpt -Chaxby -T-107/23/1 > colors.cpt` according to the data range received from GDAL. The text annotations were plotted using the `pstext` GMT module as follows (here an example for Karachi): `gmt pstext -R -J -N -O -K \ -F+f11p,Times-Roman,black+jLB -Gwhite@30 >> $ps << EOF 67.3 24.6 Karachi EOF`.

*2.2.4. Gravimetric mapping*

The most important lines of script used for plotting Faye's and Bouguer anomalies (Figs. 5 and 6) are as follows. The `img2grd` module was used to convert files from the IMG to GRD format using the following code: `gmt img2grd curv_27.1.img -R60.0/80.0/23.5/37.2 -GgravPK_B.grd -T1 -I1 -E -S0.1 -V`. The GDAL (*GDAL/OGR*, 2020) utility was used for retrieval of information by the following code: `gdalinfo gravPK_B.grd -stats`. The image was visualized by the following code: `gmt grdimage gravPK_B.grd -Cmydata.cpt -R60.0/80.0/23.5/37.2 -JM6.5i -P -I+a15+ne0.75 -Xc -K > $ps`. The contour lines were plotted using the following code: `gmt grdcontour gravPK.grd -R -J -C100 -W0.1p -O -K >> $ps`.

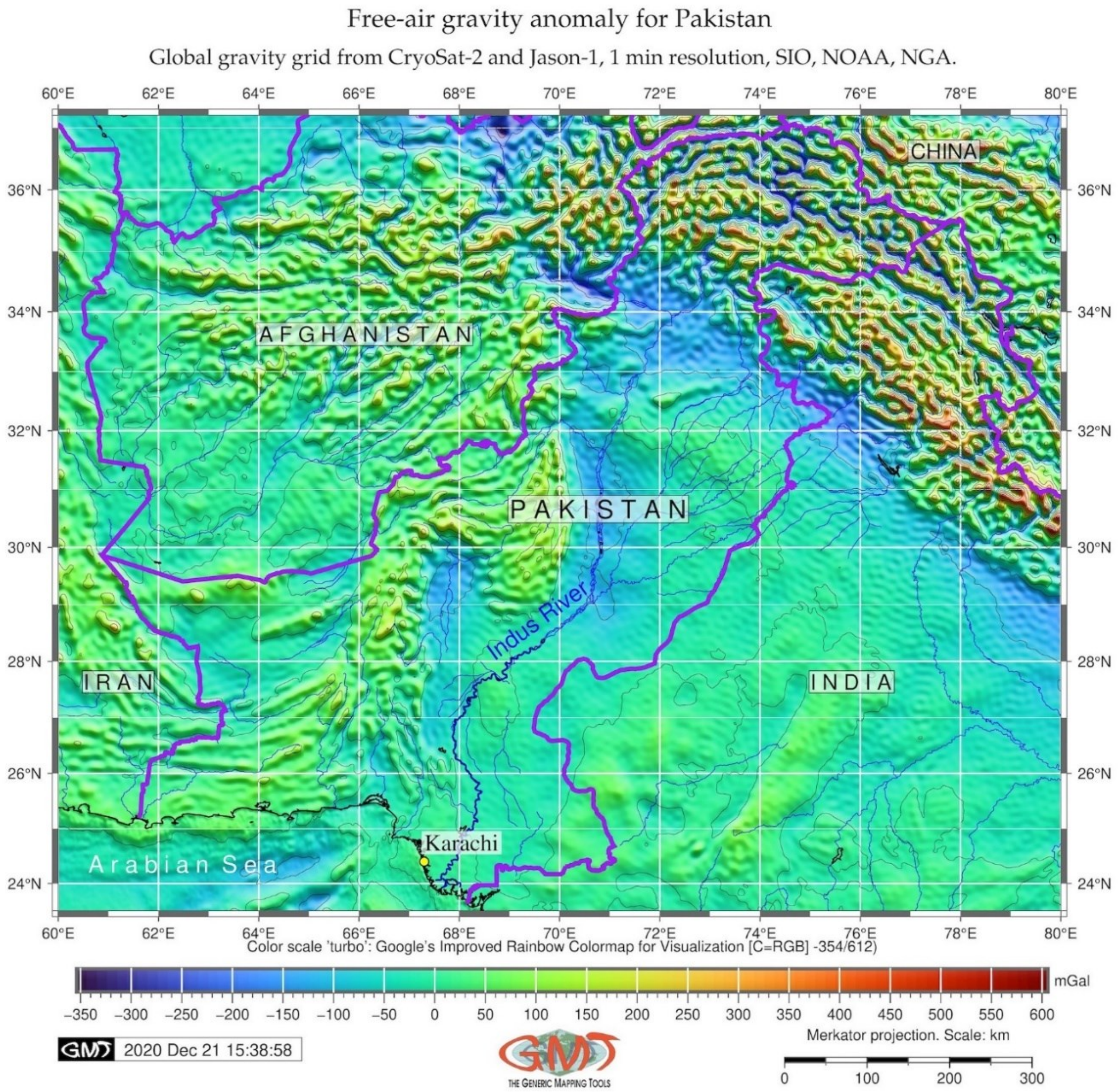

Fig. 5. Free-air gravity anomaly map of Pakistan. Mapping: GMT. Data source: geophysical anomaly grids (*Sandwell and Smith*, 1997; *Sandwell et al.*, 2014). Map production: author.

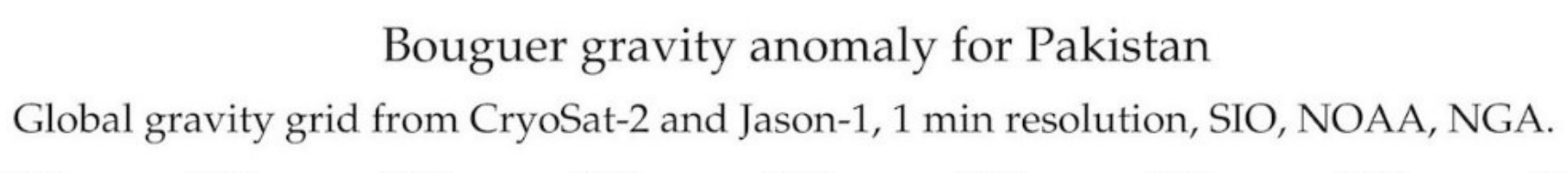

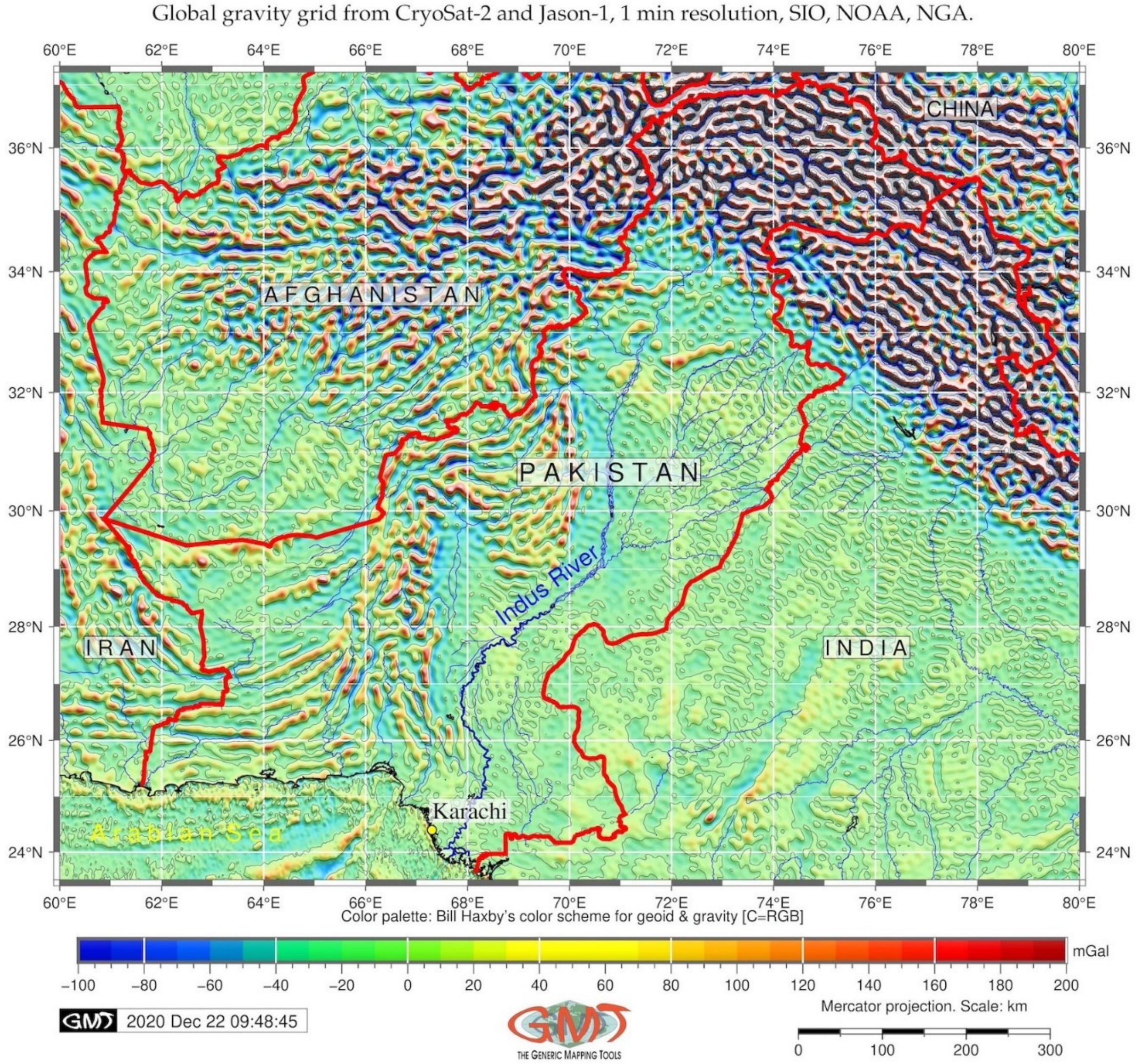

Fig. 6. Bouguer gravity anomaly map of Pakistan. Mapping: GMT. Data source: geophysical anomaly grids (*Sandwell and Smith*, 1997; *Sandwell et al.*, 2014). Map production: author.

### *2.2.5. Perspective 3D relief mapping*

The most important snippet of GMT code for 3D perspective mesh plotting is as follows: `gmt grdview pkt_relief.nc -J -R -JZ3c -Cmyocean.cpt -p115/30 -Qsm -N-2500+glightgray -Wm0.07p -Wf0.1p,red -B4/4/2000:"Bathymetry and topography (m)":ESwZ -S5 -Y5.0c -O -K >> $ps`. Here the following parameters are defined. The `grdview` module reads a 2D grid file and makes a 3D perspective plot by drawing a mesh (defined by `-Qsm` flag). The parameters of mesh lines are defined in `-Wm0.07p` (here: a very fine line thickness). The option `-N-2500+glightgray` paints a gray-shaded surface made up of polygons with a *z*-depth of −2,500, here: the shallow waters of the Arabian Sea coasts near Pakistan. The contours were plotted on top of the surface of Pakistan based on the *z*-data (elevation) information provided in a grid file. Other options include draping a data

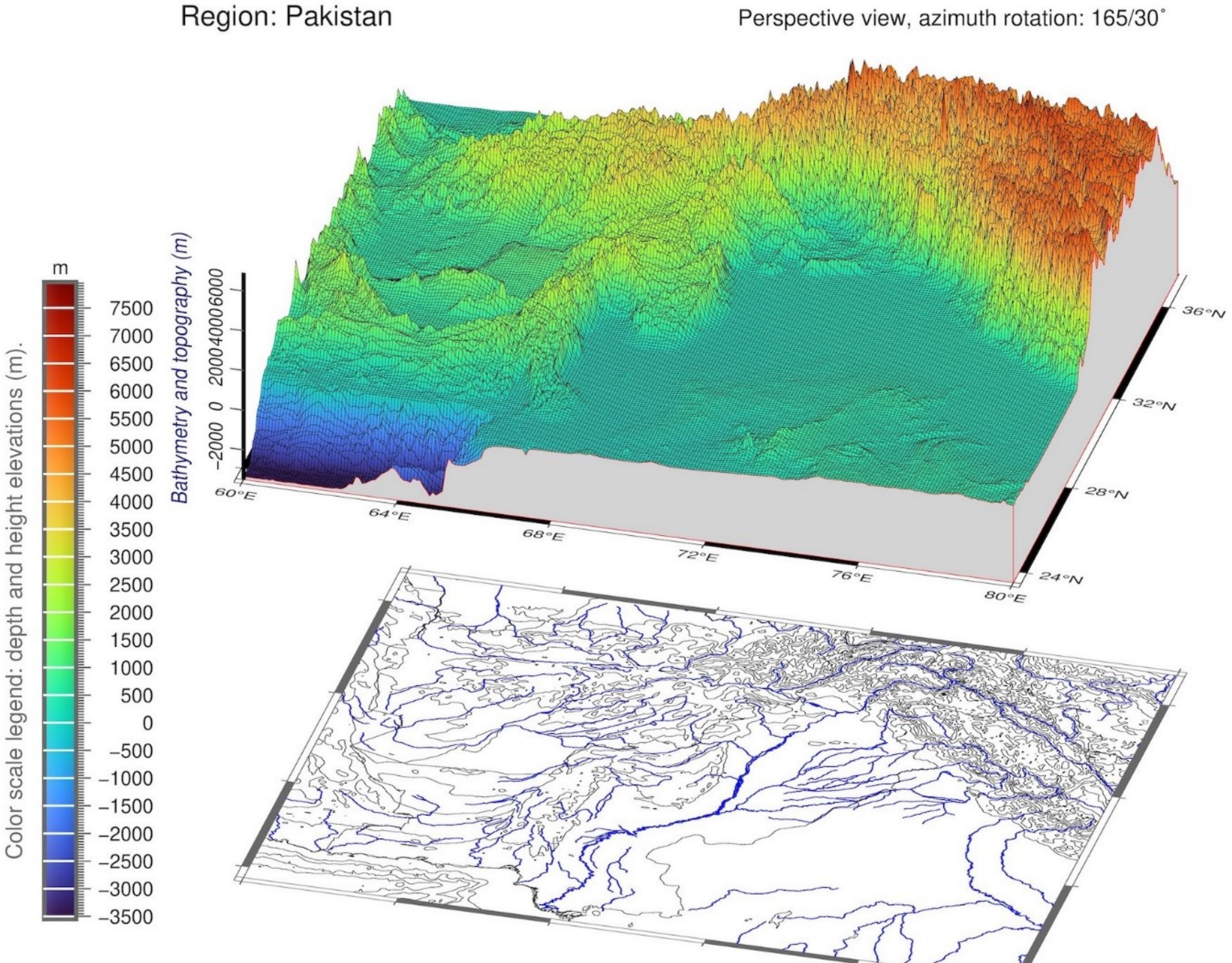

Fig. 7. 3D map of Pakistan (1). Mapping: GMT. Data source: topographic grid obtained from ETOPO5 (*National Geophysical Data Center*, 1993). Map production: author.

set on top of a surface, smoothing the contours before plotting by `S5`, setting *z*-axis scaling (`-JZ3c`), shifting plot origin (3D over the 2D) by the `-Y5.0c`. The resulting model maps are shown in Figs. 7 and 8 showing the 3D relief of Pakistan with varied rotation angles: 115° and 165°, respectively. The 2D contour maps in Figs. 7 and 8 represent ETOPO5 topographic contour in Fig. 7 and the geoid contour in Fig. 8, respectively. The EGM96 is one of the existing geoid models, commonly used in geophysical analysis (*Zou et al.*, 2015). In this study it was used to model global mean sea level of surface elevations over Pakistan.

*2.3. Scripting in R*

The part of the geospatial modelling of geomorphometric parameters of Pakistan was performed using R programming language (*R Core Team*, 2020) in the RStudio environment (*RStudio Team*, 2017). Specifically, two packages of R were used for mapping: the `tmap` (*Tennekes*, 2018) and the `raster` (*Hijmans and van Etten*, 2012).

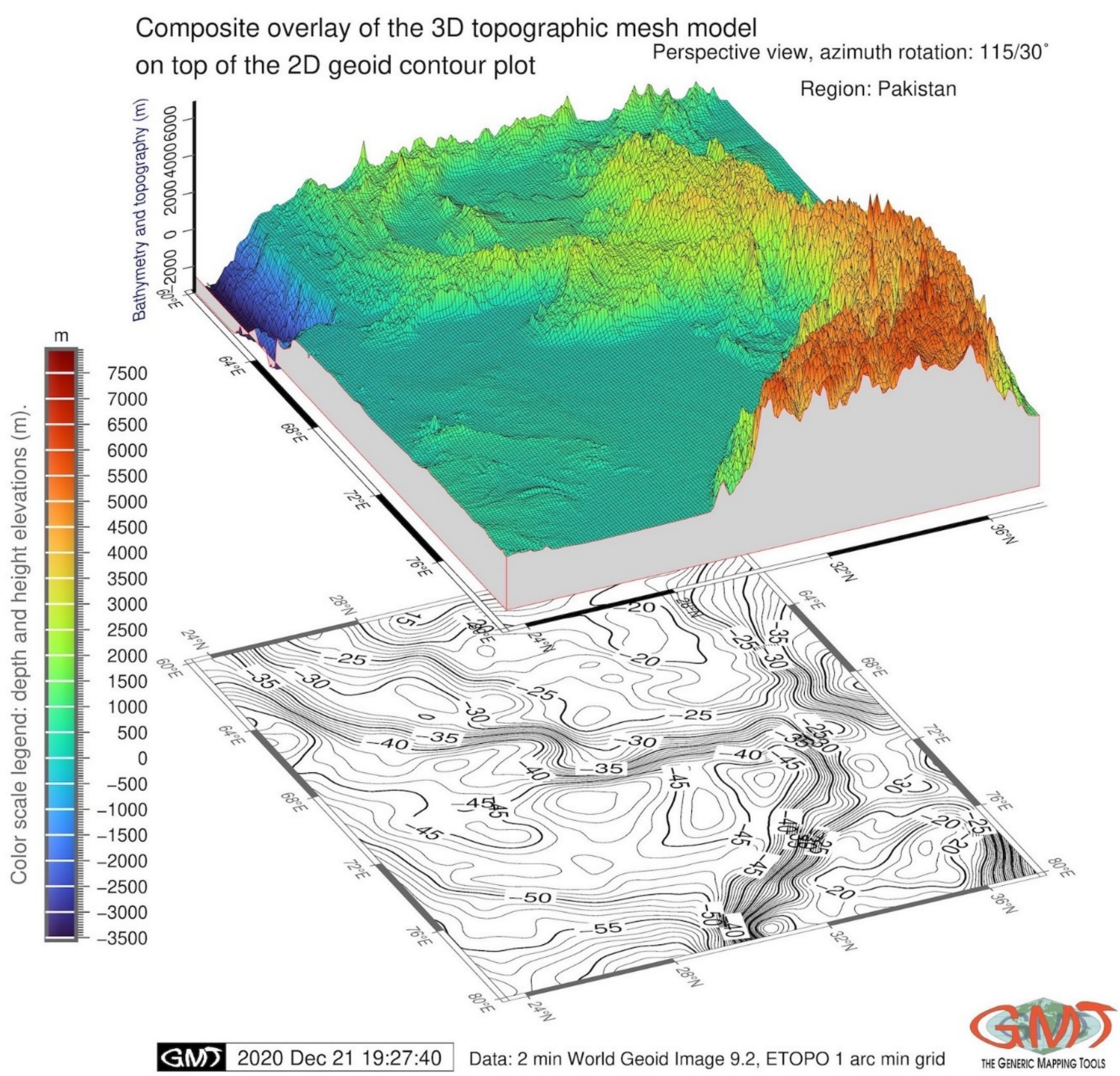

Fig. 8. 3D map of Pakistan (2). Mapping: GMT. Source: author. Data source: topographic grid obtained from ETOPO5 (*National Geophysical Data Center*, 1993) and geoid grid obtained from EGM96 (*Lemoine et al.*, 1998). Map production: author.

### *2.3.1. Slope mapping*

The present study adopts the geomorphometric methodology formulated by *Evans* (2012) and well explained in the existing works (*Sofia*, 2020; *Lemenkova*, 2021a; *Szymanowski et al.*, 2019). The process results in modelling of slope terrain criteria and yields models of aspect, slope, hillshade, and DEM that can be visualized for better highlighting the particular effects of relief and shows the area's most notable geomorphological features. Initially, a `raster` package of R was used for the study area of Pakistan using the embedded data built by compiling data from sources of the library using the following code: `alt = getData("alt", country = "Pakistan", path = tempdir())`. The three options of the package, namely `slope`, `aspect`, and `hillshade` detail the variables and the data models used in the present study using the following example: `slope = terrain(alt, opt = "slope")` which models the slope, and `plot(slope)`, which visualizes the resulting graphics using the default parameters.

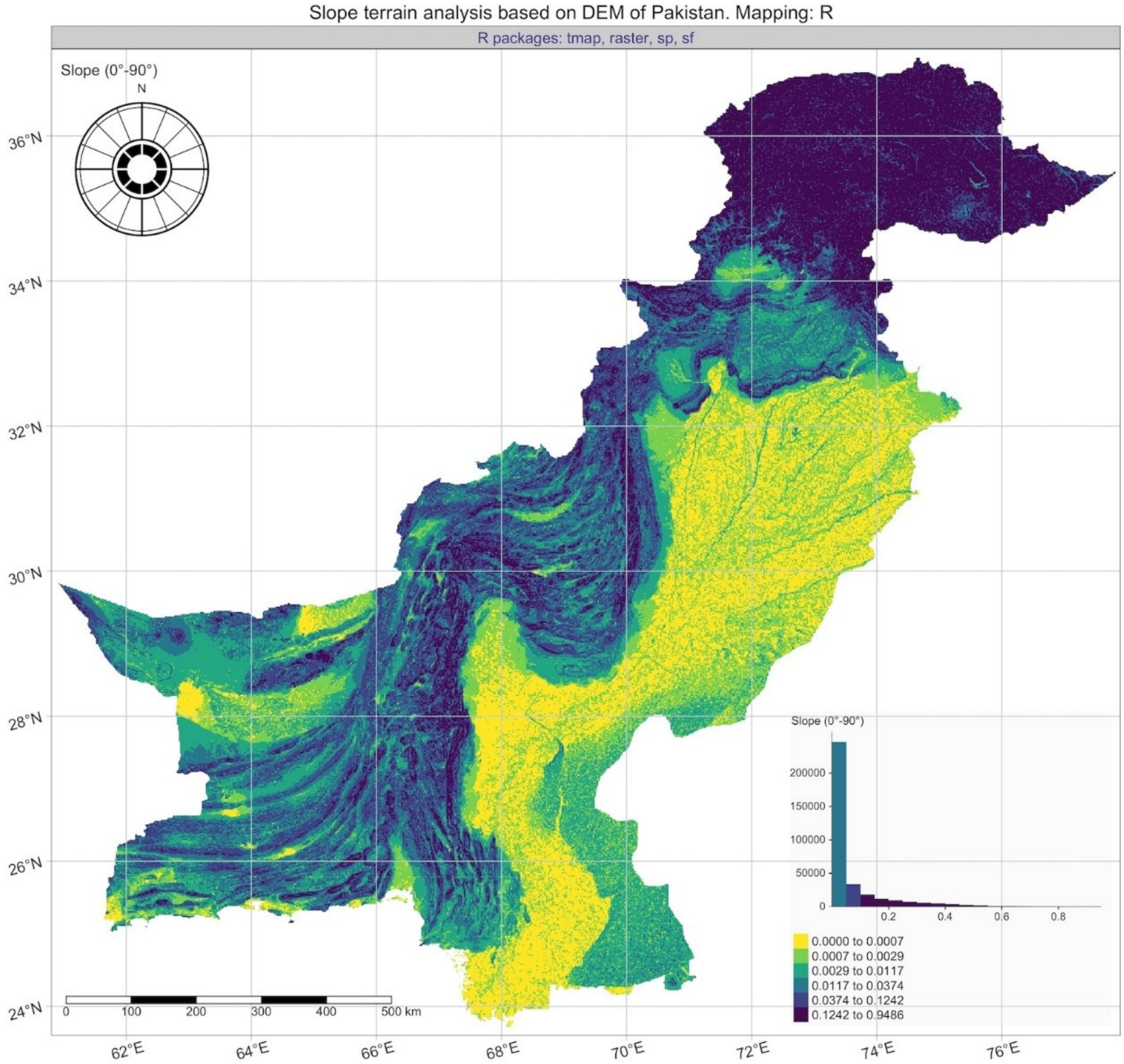

Fig. 9. Slope map of Pakistan. Mapping: R. Data source: SRTM (*NASA Shuttle Radar Topography Mission (SRTM)*, 2013; *Farr and Kobrick*, 2000; *Farr et al.*, 2007). Map production: author.

### *2.3.2. Aspect mapping*

Similar logic has been applied for aspect modelling which used the aspect modelling by the following code: `aspect = terrain(alt, opt = "aspect")`, which models the aspect, following by the `plot(aspect)`, which visualizes the aspect view by the default settings of the `raster` library of R.

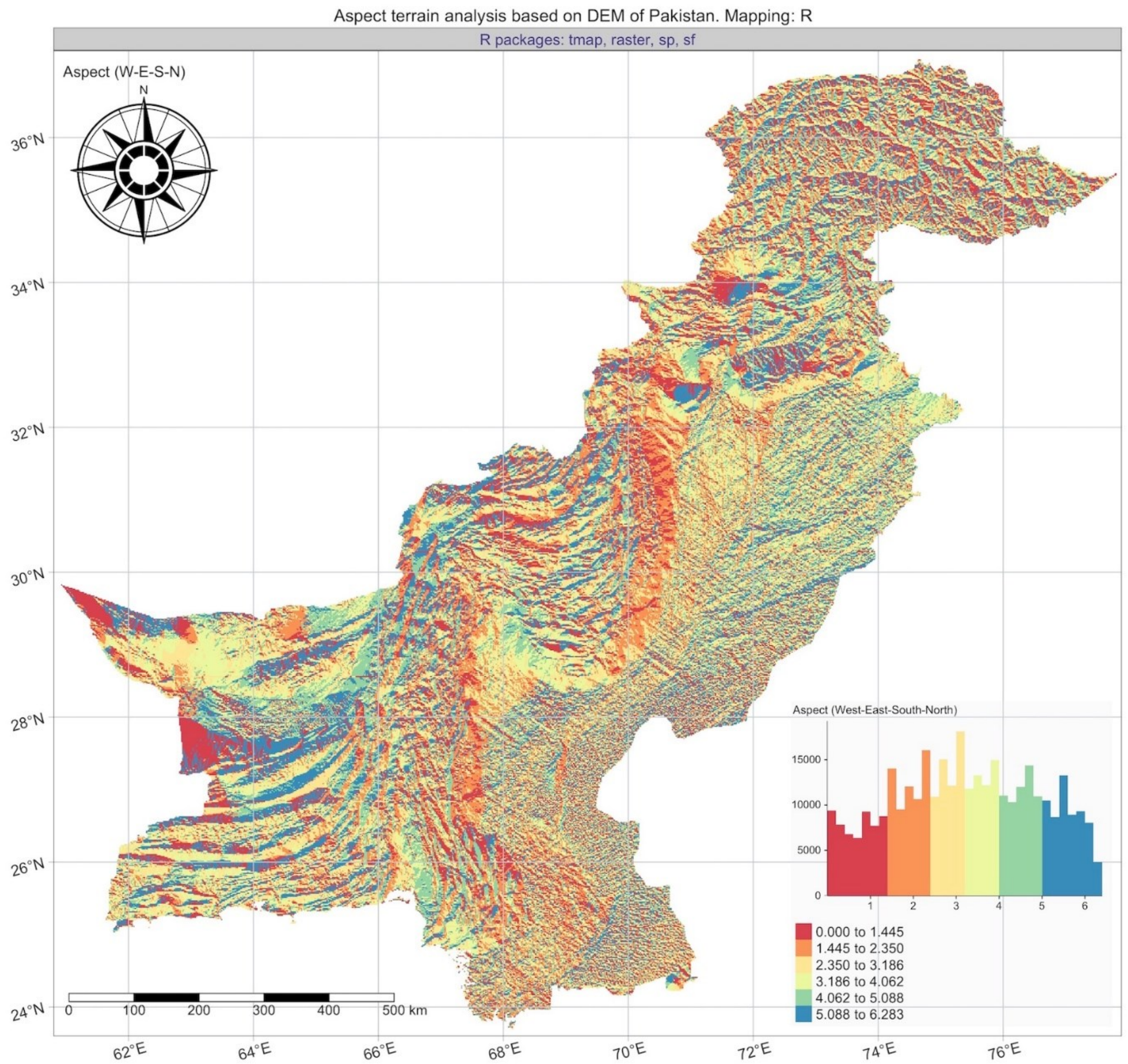

Fig. 10. Aspect map of Pakistan. Mapping: R. Data source: SRTM (*NASA Shuttle Radar Topography Mission (SRTM)*, 2013; *Farr and Kobrick*, 2000; *Farr et al.*, 2007). Map production: author.

The data values of aspect variables were assigned to the six classes showing the compass orientation of the main aspect directions of the terrain relief in Pakistan. Here a ranking was based on value ranges contributing to the rotation of the angle computed using the initial grid of DEM, while the nonnumerical visualization was made using the `Spectral` color palette of the ranked qualitatively according to the terrain properties of Pakistan geomorphology expressed in its given visualized landforms.

### *2.3.3. Hillshade modelling*

The hillshade modelling is one of the most common geomorphometric methods of relief visualization used to indicate the grayscale 3D view of the surface terrain, with the relative position of sun illumination taken into account for shading the raster graphical image. This study used the `cividis` color palette, which is generated by optimizing the well-known

`viridis` colormap to increase the effects of shading and highlights the relief of the Himalayas in the north of Pakistan and Balochistan in its south-western part.

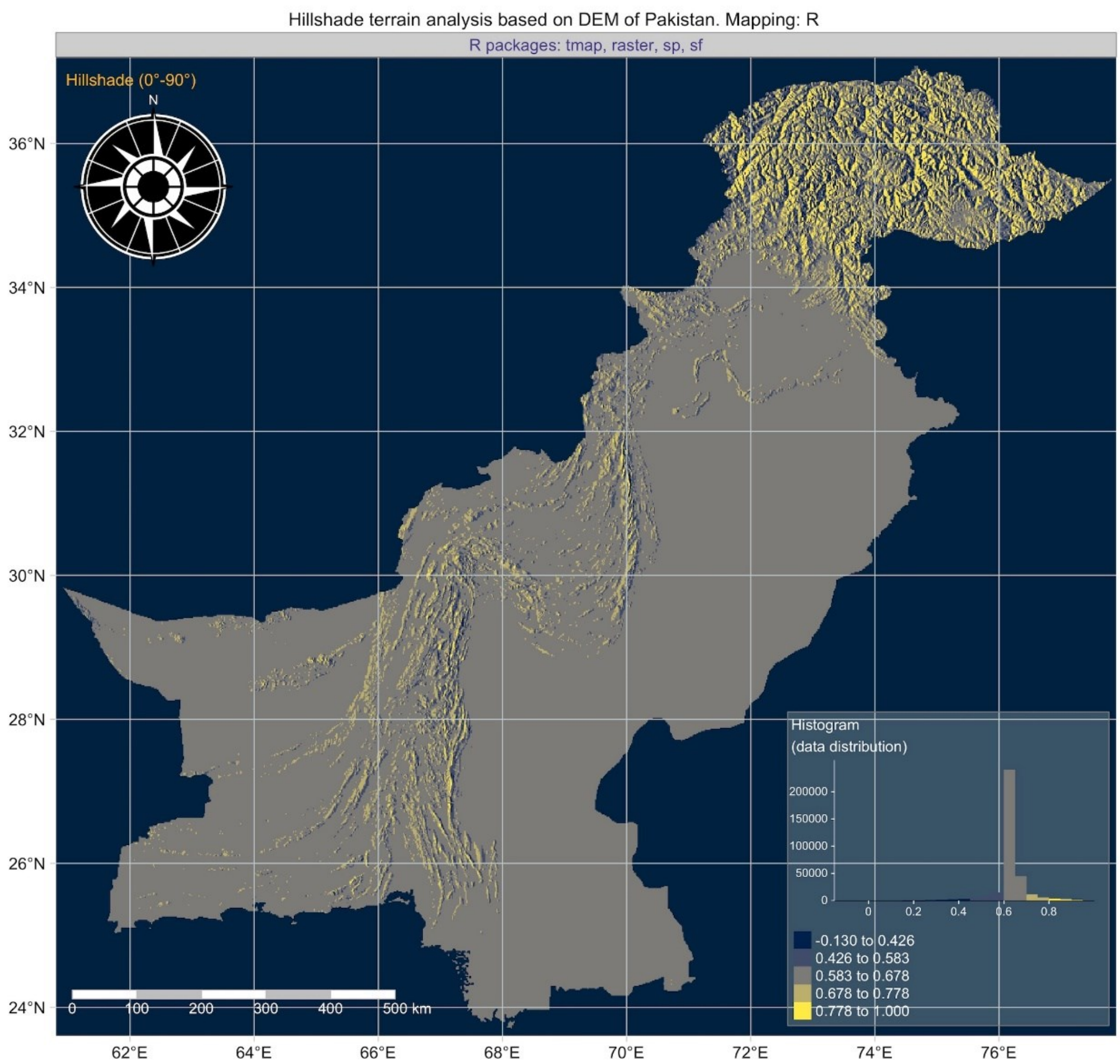

Fig. 11. Hillshade map of Pakistan. Mapping: R. Data source: SRTM (*NASA Shuttle Radar Topography Mission (SRTM)*, 2013; *Farr and Kobrick*, 2000; *Farr et al.*, 2007). Map production: author.

The hillshade is always subjected to the angle factor of the orientation which may change visually on the map due to altitude and azimuth properties to specify the position of the light source (artificial sun). These factors were controlled in the `raster` package of R by the following code: `hill = hillShade(slope, aspect, angle = 40, direction = 270)`, which states that the angle rotation was selected as 40° and the direction is 270°. From the geomorphological point of view, the slopes of Pakistan are subjected to significant variations as they differ in various regions of the country: the coastal areas, Balochistan, Sindh, Punjab, Himalaya, Muslim Bagh Ophiolites complex, which can effectively be visualized in the geomorphometric maps showing various land areas.

*2.3.4. DEM visualization*

The areas of various elevations are visualized more effectively using the `terrain.colors(256)` palette of R for mapping DEM by the `RColorBrewer` library. It is crucial to visualize regional elevation DEM in Pakistan to identify and estimate the extent of land area with varied relief types. The DEM elevation data are applied to estimate the land covered by various lithological types as shown in Fig. 2 and location of the ophiolites in Fig. 3 in response to the geologic setting of Pakistan and impacts of the tectonic movements on the distribution of mineral resources, as, for instance, described in studies on petrogenesis in Pakistan (*Khan et al.*, 2020; *Siddiqui et al.*, 2017a, 2017b). From the correlation point of view, the mountain regions with high elevation are considered as more valuable areas for mineral prospective as they provide more concentrated deposits with mineral ores and other resources against the coastal regions with low elevation values.

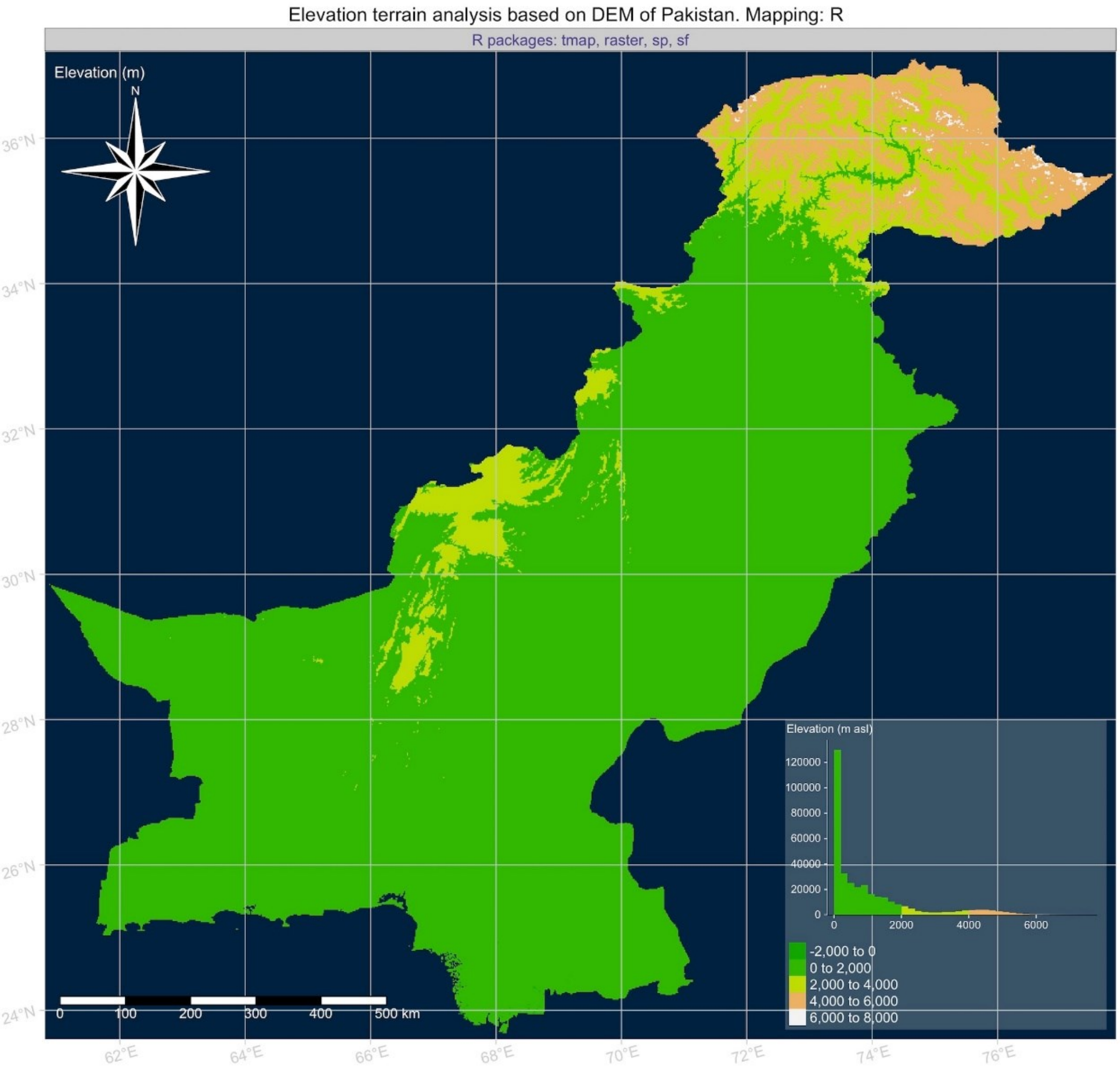

Fig. 12. DEM map of Pakistan. Mapping: R. Data source: SRTM (*NASA Shuttle Radar Topography Mission (SRTM)*, 2013; *Farr and Kobrick*, 2000; *Farr et al.*, 2007). Map production: author.

## 3 *Results and Discussion*

Pakistan is composed of four geographically and geologically distinct regions (Fig. 1): (1) Balochistan, dominated by natural resources, especially natural gas fields and notable for extremely dry desert climate; (2) Khyber Pakhtunkhwa (a.k.a., North-West Frontier Province) located on the Iranian (Persian) plateau, along the border with Afghanistan and includes one of the 14 World's highest mountains, the Nanga Parbat mountain reaching 8,126 m in its height; (3) the 3rd largest province of Pakistan, located in the south-eastern corner of Pakistan, bordering the Iranian plateau in the west (Balochistan Province) and India in the east; (4) Punjab, which is the 2$^{nd}$ largest province of the country after Balochistan, which geomorphology mostly consists of the alluvial plains of the Indus River and its tributaries. Topographically, the notable features of the Punjab Province include the Sulaiman Mountains, the Margalla Hills, the Himalaya foothills, and a hill system Salt Range (Fig. 1).

In its geologic evolution, Pakistan is formed by two distinct geological divisions: the Gondwanan and Tethyan parts. The Gondwanan segment includes the northwestern portion of the Indo-Pakistan crustal plate (the Himalayas, the Salt Range, the eastern ranges of Balochistan, the Indus Plain and the Cholistan Desert and Thar Deserts (Fig. 1).

Northern and western parts of the Indus platform have been deformed by the tectonic collision of the Indian Plate with the Eurasian Plate (Fig. 3). This caused the formation of the Large igneous provinces in the north-western part of the Indian Plate (Fig. 3). The fold-and-thrust belt is characterized by arcuate fold axes and sharp structural flexures that characterize the tectonic development of the region and orogenesis.

The quaternary sequence in Pakistan (Fig. 2, orange color for Q) represents a wide range of deposits which includes the marine deposits along the coasts of the Arabian Sea, shore and offshore deposits, the volcanic deposits of (light cyan color for Qv, Fig. 2), aeolian deposits of Thar and Cholistan deserts (light magenta color for Qs, Fig. 2), lacustrine deposits in mountain basins, deeply-weathered residual soil, glacial and fluvioglacial deposits in the Himalayas, Karakoram and Hindu Kush Ranges, as well as a series of fluvial deposits in the vast plain and delta of the Indus River (Fig. 2). The Mesozoic deposits can be seen in the northern region of Khyber Pakhtunkwa (bright magenta color in Fig. 2). The Mesozoic sedimentary rocks are deposited in Sulaiman Ranges and Salt Range areas forming significant components of the tectonic and stratigraphic sequences in the Himalayan thrust-and-fold belt and the Karakorams (Figs. 1 and 2). The Mesozoic rocks form a thick, buried cover sequence over the basement rocks in most segments of the Indian Plate (Fig. 3).

The Sulaiman fold-and-thrust belt, situated in the northern part of Balochistan province, and partially in southwestern Khyber Pakhtunkhwa, is a remarkable geologic region of Pakistan. Having a notable arcuate geometric form, it consists of Mesozoic shelf carbonates, shales and volcanics, Paleogene shallow-water marine and continental deposits and Neogene molasse. It is characterized by the E–W tending crescent-formed, convex southward folds bending northwards on both sides and comprised of Jurassic to Recent sedimentary rocks (Figs. 1, 2, and 3). On the east, the folds gradually change direction into the N–S directed folds of the Sulaiman Range. The Paleogenic rocks (Pg, khaki color in Fig. 2) are notable in the Sulaiman fold-and-thrust belt which present a complex arcuate alternation between the rocks from the Paleogene (lower Eocene) formation with the Middle Eocene (Kirthar) and Neogene (recent alluvial sediments, light rose in Fig. 2). As recently found (*Jadoon*, 2019), the basal decollement is situated in the Paleozoic pelitic carbonates at a depth of about 10 km of the Sulaiman Belt. The Sulaiman Fold Thrust Belt is tectonically active and notable for the compressional deformation, uplift, and erosion (*Khan*, 2019), which affected the shaping of structural and depositional architecture of the basin.

A special part of the Sulaiman Fold Belt is presented by the Quetta–Muslim Bagh–Sibi Region where a Muslim Bagh ophiolite formation was formed in with the Spongtang Arc. The chain of Muslim Bagh ophiolites within western Pakistan was formed in the obduction zone on the Indian passive margin, which formed an arc system formed as the oceanic crust in Tethys during India migration to the north. The Muslim Bagh present the remnants of the oceanic lithosphere tectonically replaced on continental margins as a result of complex geologic evolution, incorporated in island arcs, and then deformed and altered blocks in the collision belts. Therefore, the ophiolites have key importance in the geological history of the zones where they are located. The nature and origin of the lithological units of the ophiolites enable the model tectonic evolution of the geological province, pointing at their origin to their location of the ophiolite complex. The notable ophiolite complexes in Pakistan are located along the northern and western borders of the Indian Plate along the Afghan block. Their larger complexes (Bela and Zhob valley ophiolites) are situated along the western suture in Balochistan with the Zhob ophiolites consisting of Muslim Bagh Khanozai and Zhob ophiolites, consequently.

The presented map of the geoid model based on the EGM-2008 (Fig. 4) shows a remarkable correlation in isolines with the stretching of the Karakoram complex of a narrow belt of metamorphic rocks which lies to the east of Nanga Parbat. The geoid undulations can be explained by the crust and upper mantle variations reflecting the density and porosity in rocks. The clearly visible difference in geoid heights (Fig. 4) shows variations in crustal structure between Afghanistan (−15 to 20 m, Fig. 4) and India with negative geopotential values (below −65 m, cyan to blue colors in Fig. 4). the situation in Pakistan is rather varied and complex with notable heights corresponding to the Suleiman Thrust Block, the northern mountainous territory of the Khyber Pakhtunkhwa (the Himalayas and the Karakorum) and contrasting Lower Indus Basin (Sindh Province), Middle Indus Basin and Upper Indus Basin (Punjab Province) where the geoid heights mostly lie within the range of −75 to −55 m. The geoid undulations and variations over Pakistan could be related to rock density inhomogeneities in the upper mantle resulted from the processes of rock formation.

The anomalies in the geophysical fields (Faye's and Bouguer) of the study area are presented in Fig. 5 and 6, respectively for Faye's and Bouguer gravity anomaly maps. The importance of the gravity anomaly modelling consists in the additional information that can be derived from the data of satellite missions. The global and regional constraints on the gravity fields enable to better analyze geophysical processes and dynamics (*Poutanen et al.*, 2009). The slightly positive correlations of the free-air gravity anomalies with positive values of 0 to 100 mGal (light aquamarine colors) are visible for the area of the Sulaiman Fold Belt, and northern regions of Pakistan which corresponds to the granite suites of Himalaya, Karakorum and the Hindu Kush reaching its highest values in Karakorum range (orange to red colors in Fig. 5). Not much principal difference was detected between the observed Faye's and Bouguer gravity anomaly maps (Figs. 5 and 6) in a comparative analysis since the maps correspond to the geophysical setting of the region with the difference in the algorithm. Hence, the Bouguer gravity anomaly represents the one corrected for the terrain height at which it was measured and the attraction of terrain, while height correction alone gives a free-air gravity anomaly.

The analysis of the 3D maps (Figs. 7 and 8) presents the perspective mesh plots visualizing the 3D gridding of the terrain area of Pakistan with varied azimuth rotations as of 165/30° and 115/30° for Figs. 7 and 8, respectively. The composite overlay in Fig. 8 shows the correlation of the terrain with the geoid isolines (shown as a 2D subplot map) that proves the impact of the density and porosity of the geological substrata in the underlying rocks on the geoid undulations. The arcuate view of the geoid isolines depicts the line of the Sulaiman Fold Thrust Belt continuing to the Hindukush, Pamir, Himalaya and Karakorum ranges as can be seen on the 3D view. Likewise, Fig. 7 gives us the perspective analytical view of the well-defined Indus River basin well contrasting with the adjusting mountain ranges and Sulaiman Fold Belt with a comparative view of the tributaries of Indus: Lower Indus Basin (Sindh), Middle Indus Basin and Upper Indus Basin (Punjab). The geomorphological analysis (Figs. 9, 10, 11, and 12) presents the four types of geomorphometric mapping that visualize the relief terrain of the country based on DEM. The slope classification in Fig. 9 shows the 6 classes divided by the slope steepness.

## *4 Conclusion*

In this paper, the endeavors were made to integrate and summarize the available geologic and geophysical information using the available open source tools. The lithological data have been visualized using QGIS. The geophysical datasets were used for mapping free-air gravity anomaly of Faye's and Bougier. The data include the geoid EGM-2008 grid which shows inter-spatial correlation with geological units within the country of Pakistan. The tectonic events, the extension of tectonic slabs and plates have been incorporated in the GMT and data visualized by various GMT modules. The geomorphometric units (slope, aspect, hillshade and DEM elevation) have been mapped using R libraries `raster` and cartographically adjuster by `tmap`. Efforts have been made to incorporate the ETOPO raster grids for 3D modelling of the relief terrain of Pakistan with changed rotation (115° and 165°).

Following general conclusions can be drawn from this research.

(i) In the study area of Pakistan, the distribution of the geological structural elements shows that the geologic history and evolution of Pakistan affected significantly the geophysical and geomorphological setting of the country that is notable in the comparison of the thematic maps.

(ii) The distribution of the relief varies according to the types of the tectonic formation in northern (Himalaya, Karakorum, Pamir), central (Sindh), western (Balochistan) and eastern (Punjab) parts.

(iii) The ophiolite are distributed very uneven with clearly visible Muslim Bagh ophiolite complex, belts of the ophiolites located in the Himalaya, Karakoram and Pamir mainly in northern mountainous regions.

(iv) The geophysical anomaly fields, including the Bouguer and Faye's gravity anomaly fields, show gradual changes in absolute values that reflect the rock inner properties of the underlying geologic layers in all regions.

(v) The 3D models showing the relief variations present correlation with geoid undulation isolines and valley of the Indus River (Lower Indus Basin in Sindh, Middle Indus Basin and Upper Indus Basin in Punjab) well corresponds with the fluvial network that contributes to the sediment deposits in the valley.

(vi) The geomorphometric analysis demonstrated the variations in the aspect and slope steepness of the country's topography and visualization of the hillshade and elevation based on DEM using R language.

The novelty of the presented research consists in the significance of the study area encompassing the whole country of Pakistan with an accent in Sulaiman Fold Belt and Muslim Bagh ophiolite complex segments. Compared to other publications (*Farid et al.*, 2013; *Muhammad et al.*, 2019; *Hakro et al.*, 2016; *Kasi et al.*, 2014), this research presents the integrated study both for the country-level in Pakistan as the study area and for the regional level indicating locations of the geologically special parts of the country, which has a much larger area than the previously published articles and thus, well contributes to the existing works on the geology of Pakistan (*Kakar et al.*, 2015; *Usman et al.*, 2020; *Siddiqui et al.*, 2015). Besides the above said, the application of the scripting methods, presented in detail with practical examples of the GMT and R codes, and the use of the free open source geophysical, geological, geomorphological and topographic data that were used in this investigation make this research repeatable and maps useful in similar research. In this way, the present research serves as a valuable contribution to the studies on the geology of Pakistan.

*Acknowledgement*

The author is grateful to the reviewer Prof. Markku Poutanen from the Finnish Geospatial Research Institute (FGI) for review, comments, and editing of this manuscript.